\documentclass[epj]{svjour}
\usepackage{graphicx}
\usepackage{amsmath}
\usepackage{amssymb}
\usepackage{color}
\begin{document}
\title{Study of implosion in an attractive Bose-Einstein condensate}
\author{S. Sabari\inst{1} Amitava Choudhuri\inst{2} K. Porsezian\inst{3} Bishwajyoti Dey\inst{1}}

\authorrunning{S. Sabari et al.}
\titlerunning{Implosion of an attractive BEC}
\institute{Department of Physics, University of Pune, Pune 411007, Maharashtra, India \and Department of Physics, The University of Burdwan, Golapbag, Burdwan 713104, West Bengal, India \and Department of Physics, Pondicherry University, Puducherry 605014, Puducherry, India}

\abstract{By solving the Gross-Pitaevskii equation analytically and numerically, we reexamine the implosion phenomena that occur beyond the critical value of the number of atoms of an attractive Bose-Einstein condensate (BEC) with cigar-shape trapping geometry. We theoretically calculate the critical number of atoms in the condensate by using Ritz's variational optimization technique and investigate the stability and collapse dynamics of the attractive BEC by numerically solving the time dependent Gross-Pitavskii equation.
 % 67.85.Hj, 67.85.Jk, 03.75.Kk end of PACS codes
} %end of abstract
\maketitle

\section{Introduction}
\label{sec1}

Since the experimental realization of Bose-Einstein condensates (BECs) two decades ago \cite{Anderson1995,Davis1995,Bradley1995}, the field of quantum gases has attracted much theoretical and experimental interest. At absolute zero temperature the properties of a condensate are well described by the nonlinear, mean-field Gross-Pitaevskii equation (GPE) for both repulsive and attractive inter-atomic interactions, see the early comprehensive reviews  \cite{Dalfovo1999,Leggett2001,Kevrekidis2004,Abdullaev2005}; for more recent overviews in the area of ultracold quantum Bose and Fermi gases and related phenomena, see, for instance, Refs.  \cite{Giorgini2008}.  The nonlinear terms in the GPE arise from the interactions between atoms in the condensate, characterized by their two-body s-wave scattering length, $a_s$. The static and dynamical properties of trapped BEC crucially depend on the sign and strength of the interatomic interactions. Both the strength and the sign of the interaction can be controlled by varying the scattering length $a_s$ of atoms near the Feshbach resonance \cite{FBR}. The stability of the condensate under magnetic traps has been studied  both numerically and analytically \cite{Ruprecht,Dalfovo,attractive}. If the interaction is repulsive, the BEC is stable and its size and number of particles have no fundamental limit. If it is attractive, only a limited number of atoms can form a condensate. Moreover, the BEC was predicted to be metastable, which means that it remains stable for some finite time, only when the number of atoms is below some critical number ($N_c$). The estimated critical number for $^7$Li atoms is about $10^3$ \cite{Bradley1997}. Beyond  $N_c$, we expect the BEC to `collapse' or `implode' due to the strong attactive interaction. In particular, in a trapped attractive gas whose number of particles or the strength of the inter-particle interaction exceeds a critical value, the kinetic energy cannot balance the (negative) interaction energy \cite{PRA83,Shuryak}. The attraction brings the bosons so tightly close such that the spatial extension of the wave function of the system shrinks to a point and the condensate eventually implodes \cite{PRA83,PhysRevA82}. A BEC can avoid implosion only as long as the number of atoms in it, $N$, is less than a critical value $N_c$. Beyond $N_c$, we show in our theoretical study that if we increase the number of atoms, the BEC first explodes, then implodes as expected, and subsequently a sudden explosion of atoms occurs. The explosion followed by an implosion has been already studied experimentally.
This `explosion' actually corresponds to a small amount of energy by normal standards. The explosion of a collapsing BEC was termed as `bosenova' \cite{Donley2001}.

The study of collapsing phenomena of attractive BECs has been the subject of much interest for the past two decades; see, for instance, Refs. \cite{Ruprecht,Dalfovo,attractive,Bradley1997,PRA83,Shuryak,PhysRevA82,Donley2001,Melo,talukdar,HU2000}. It refers to the situation when strong self-focusing of a beam leads to a catastrophic increase (blow-up) of its intensity in a finite time or after a finite propagation distance. In this context, the collapse process is a fast, collective phenomenon consisting of the destruction of a multi-particle system happening abruptly on the timescale that governs the usual dynamics. One example is the collapse of the gravitational core, initiating a supernova. Tailoring both the external confining potential and the interaction between the atoms allows us to control the properties of the condensate. It is an ideal system to study not only problems of condensed matter physics, but also the dynamics of a collapse as well. The collapse of an attractive BEC of $^7{Li}$ or $^{85}Rb$ atoms has been observed by various experimental groups \cite{Bradley1997,roberts}(collapsing condensates were first observed in $^7$Li \cite{NJP}) and theoretically analyzed by many authors for various external trapping potentials such as, single-well potential \cite{Ruprecht,Sackett}, potential without axial confinement \cite{Garcia,Carr}, toroidal confinement \cite{Parola}, double-well potential \cite{Sakellari}, and periodic potential \cite{Adhikari}. In recent past, Ghosh \cite{Ghosh} studied the duality-symmetry, which are abundant in different branches of physics and astrophysics (see \cite{Ghosh} and refs. therein) and showed that invariance under the
duality-symmetry leads to explosion-implosion duality in one- and two-dimensional Bose-Einstein condensation without the harmonic trap.
\par
In this work, we study the sequence of implosion and explosion phenomena of an attractive BEC beyond the $N_c$, by using the time-dependent, nonlinear, mean-field GPE. First, we calculate theoretically the $N_c$ for the BEC by solving the quasi one-dimensional (Q1D) GPE taking recourse to the use of Ritz's variational optimization technique \cite{Anderson1983}. Then, we study the collapsing phenomenon of an attractive BEC beyond $N_c$ through extensive numerical simulations of the GPE. The organization of the present paper is as follows. In Sec. \ref{sec2}, we present a brief overview of the nonlinear mean-field model. In Sec. \ref{sec3}, we derive the equation of motion of the Q1D system to find out the critical number of atoms through variational approximation (VA) method and discuss the stability of the condensate. Then, in Sec. \ref{sec4} we numerically study the collapsing dynamics of the BEC by analyzing the GPE through split-step Crank-Nicholson (SSCN) method \cite{Muruganandam2009}. Finally, we give the concluding remarks in Sec. \ref{sec5}.

\section{Nonlinear mean-field model}
\label{sec2}

The GPE can be used at low temperatures, to explore the macroscopic behavior of the system. The time-dependent BEC wave function $\Psi(\tilde{\textbf{r}},\tau)$ can be described by the following mean-field nonlinear GP equation \cite{Dalfovo1999},
\begin{eqnarray}
\left[{-i \hbar \frac{\partial}{\partial t}}-\frac{\hbar^2}{2m}\nabla^2+V(\tilde{\textbf{r}})+g N|\Psi(\tilde{\textbf{r}},\tau)|^2\right]  \Psi(\tilde{\textbf{r}},\tau) = 0. \label{Jac1}
\end{eqnarray}
Here $\nabla^2=\frac{\partial^2}{\partial x^2}+\frac{\partial^2}{\partial y^2}+\frac{\partial^2}{\partial z^2}$, $g=4\pi \hbar^2a_s/m$, $a_s$ is the atomic s-wave scattering length, which is negative or positive for attractive or repulsive interaction between atoms in the condensate, $m$ is the mass of a single bosonic atom and $N$ is the number of atoms in the condensate. The external magnetic trap potential may be written as $V(r)=\frac{1}{2}m\omega^2(\nu^2x^2+\kappa^2y^2+\lambda^2 z^2)$, where $\omega_x\equiv \nu / \omega$, $\omega_y \equiv \kappa /\omega$ and $\omega_z \equiv \lambda/ \omega$ are the angular frequencies of the trap in the $x,y$, and $z$ directions, respectively. The normalization condition is $\int |\Psi(\tilde{\textbf{r}},\tau)|^2 = 1$.

To study the collapse dynamics beyond the critical number of atoms, we are interested to work with quasi one-dimensional (Q1D) cigar-shaped  BEC dispersing along the $z$ direction. In the following, we consider equation (\ref{Jac1}) in a geometry in which the trapping potential in $z$ is much weaker than the corresponding potential in $\tilde{\rho} = \sqrt{(\tilde{x}^2 + \tilde{y}^2)}$. Further, we make the transformation of variables as $\rho = \sqrt{2}\tilde{\tilde{\rho}}/l$, $z = \sqrt{2}\tilde{z}/l$, $t = \tau\omega$, $l = \sqrt{\hbar/(m\omega)}$ and  $\phi(\rho,z,t)=\Psi(\tilde{\rho},\tilde{z},\tau)(l^3/2)^{1/2}$. Then, the GP equation (\ref{Jac1}) becomes \cite{Muruganandam2009},
\begin{eqnarray}
\left[{-i\frac{\partial}{\partial t}}-\nabla^2+\frac{1}{4}(\rho^2+\lambda^2z^2)+{ \frac{8\pi N a_s}{l}\left\vert\phi\right\vert^2} \right]\phi = 0, \label{Jac1a}
\end{eqnarray}
We assume a separable ansatz for the solution of equation (\ref{Jac1a}) such that \cite{Garcia}
\begin{eqnarray}
\phi(\rho,z,t)=u(\rho)\psi(z,t)\,\,. \label{Jac1b}
\end{eqnarray}
Substituting Eq. (\ref{Jac1b}) in Eq. (\ref{Jac1a}), after simplification one can get the following Q1D equation
 \begin{eqnarray}
{i\frac{\partial \psi}{\partial t}}+\frac{\partial^2\psi}{\partial z^2}-\lambda^2z^2\psi-\frac{4\pi N a_s}{l}|\psi|^2  \psi = 0, \label{Jac2}
\end{eqnarray}
with
\begin{eqnarray}
\int^{\infty}_{-\infty}|\psi|^2 dz= N/\pi. \label{Jac2a}
\end{eqnarray}
Equation (\ref{Jac2}) represents a desired form of the evolution equation in which the atom-atom interaction is characterized by a negative s-wave scattering length. The realistic 1D limit in (\ref{Jac2}) is not a true 1D system because this equation involves the effect of transverse degrees of freedom through $z$ and $l$.

\section{Variational calculation of critical number $N_c$}
\label{sec3}

In this section by using  Ritz's variational optimization technique \cite{Anderson1983}, we shall calculate the critical number of atoms ($N_c$) to analyze the collapsing dynamics of the matter waves beyond $N_c$. The Eq. (\ref{Jac2}) can be restated as the following variational problem
\begin{eqnarray}
\delta\int\mathcal{L}(z,\,t,\,\psi,\, \psi^*,\,\psi_z,\,\psi^*_z,\,\psi_t,\,\psi^*_t)dt=0. \label{Jac3}
\end{eqnarray}
with the Lagrangian density written as
\begin{eqnarray}
\mathcal{L}(t) = &\,  \frac{i}{2} \left(\psi_t^* \psi - \psi_t \psi^*   \right)+\left\vert \frac{\partial\psi}{\partial z} \right\vert^2 +\lambda^2z^2\vert  \psi\vert^2+\frac{g}{2} \vert  \psi\vert^4. \label{Jac4}
\end{eqnarray}

Now we use the variational approach with the Gaussian trial wave function for the solution of Eq. (\ref{Jac2}) \cite{Garcia,Sabari2010}:
\begin{eqnarray}
\psi(z,t) = A(t)\exp{\left[-\frac{z^2}{2R(t)^2}+\frac{i}{2} \beta(t)z^2\right]}, \label{Jac4a}
\end{eqnarray}
where $A(t)$, $R(t)$,  and $\beta(t)$ are the time dependent amplitude, width, and chirp, respectively. The initial condensate at rest will have $dR(t)/dt = 0$. The trial wave function equation (\ref{Jac4a}) is substituted in the Lagrangian density and the averaged effective Lagrangian is calculated by integrating the Lagrangian density as $L_{eff} = \int^\infty_\infty \mathcal{L}\, dz$ to write
\begin{eqnarray}
 <\mathcal{L}(t)>&\,=\sqrt(\pi)\left[\frac{i}{2}(\psi\psi_t^*-\psi_t\psi^*)R\right]+\sqrt(\pi)
 \dot{\beta}R^3 \vert\psi\vert^2 \nonumber\\
 &\,+ \sqrt(\pi) g\vert\psi\vert^4 + \frac{\sqrt(\pi)}{2} (\lambda^2 R^3 + \beta^2 R^3 + \frac{1}{R}\vert\psi\vert^2)
 &\, \label{Jac4b}
\end{eqnarray}
The reduced variational principle
\begin{eqnarray}
\delta\int<\mathcal{L}>dt=0. \label{Jac4c}
\end{eqnarray}
gives a set of coupled ordinary differential equations for the parameters of our trial solution. Let us now obtain the variational equations for the Gaussian parameters $A(t)$, $A(t)^*$, $R(t)$, and $\beta(t)$, which follow from the vanishing conditions of $\frac{\partial  <\mathcal{L}>}{\partial A^*} $, $\frac{\partial  <\mathcal{L}>}{\partial A} $, $\frac{\partial  <\mathcal{L}>}{\partial R} $, and $\frac{\partial  <\mathcal{L}>}{\partial \beta} $. These equations are given by
\begin{eqnarray}
\frac{\partial  <\mathcal{L}>}{\partial A^*} = &\,\frac{1}{2 R}A+\frac{1}{2} A R^3 \beta^2+\frac{1}{2} A R^3 \lambda^2+\frac{1}{\sqrt{2} l} A^2 \pi  R a_s A^*\nonumber\\
 &\,-i R \dot{A}-\frac{1}{2} i A \dot{R}+\frac{1}{4} A R^3 \dot{\beta} = 0
\label{Jac5a}
\end{eqnarray}
\begin{eqnarray}
\frac{\partial  <\mathcal{L}>}{\partial A} = &\,\frac{A^*}{2 R}+\frac{1}{2} R^3 \beta ^2 A^*+\frac{1}{2} R^3 \lambda ^2 A^*+\frac{A \pi  R a_s \left(A^*\right)^2}{\sqrt{2} l}\nonumber\\
 &\,+\frac{1}{2} i A^* \dot{R}+\frac{1}{4} R^3 A^* \dot{\beta}+i R \dot{A^*} = 0
\label{Jac5b}
\end{eqnarray}
\begin{eqnarray}
\frac{\partial  <\mathcal{L}>}{\partial R} = &\,-\frac{A A^*}{2 R^2}+\frac{3}{2} A R^2 \beta ^2 A^*+\frac{3}{2} A R^2 \lambda ^2 A^*-\frac{1}{2} i A^* \dot{A}\nonumber\\ 
&\,+\frac{A^2 \pi  a_s \left(A^*\right)^2}{2 \sqrt{2}l}+\frac{3}{4} A R^2 A^* \dot{\beta}+\frac{1}{2} i A \dot{A^*} = 0
\label{Jac5c}
\end{eqnarray}
and
\begin{eqnarray}
\frac{\partial  <\mathcal{L}>}{\partial \beta} = &\,A R^3 \beta  A^*-\frac{1}{4} R^3 A^* \dot{A}-\frac{3}{4} A R^2 A^* \dot{R}\nonumber\\ 
&\,-\frac{1}{4} A R^3 \dot{A^*} = 0.
\label{Jac5d}
\end{eqnarray}
From equations (\ref{Jac5a}) and (\ref{Jac5b}) we get
\begin{eqnarray}
\frac{\partial}{\partial t} (RAA^*) & = & 0 \label{Jac6}
\end{eqnarray}
and consequently
\begin{eqnarray}
R\vert A \vert^2 & = & Q,  \label{Jac7}
\end{eqnarray}
where the constant $Q$ is  related to the number of particles in the condensate since the value of the integral (\ref{Jac2a}) is $\sqrt{\pi}R\vert A \vert^2$. Combining the equations (\ref{Jac5d}) and (\ref{Jac7}), we get
\begin{eqnarray}
\beta & = & \frac{1}{2}\frac{d}{d t}(ln \, R). \label{Jac8}
\end{eqnarray}
Equations (\ref{Jac7}) and (\ref{Jac8}) clearly show that if we can derive a method to calculate the values of $R$, the other parameters of the condensate will be automatically determined. It is fortunate that we are able to write a second-order ordinary differential equation from (\ref{Jac5a}), (\ref{Jac5b}), (\ref{Jac5c}), and (\ref{Jac8}). This gives a first integral of the form
\begin{eqnarray}
\frac{1}{2}\left(\frac{d R}{d t}\right)^2+2\lambda^2 R^2+\frac{\sqrt{2}\pi N a_s}{l}\frac{1}{R} +\frac{2}{R^2}& = & E, \label{Jac9}
\end{eqnarray}
where $E$ is the the constant of integration. The equation for $R$ in Eq. (\ref{Jac9}) is related to the motion of a particle of unit mass in a potential field $V(R)$ of the form
\begin{eqnarray}
V(R)& = & \frac{2}{R^2}+ 2\lambda^2 R^2+\frac{P}{R}\,\,,\,\,\,\, P = \frac{\sqrt{2}\pi N a_s}{l}.\label{Jac10}
\end{eqnarray}
The constant of the motion $E$, i.e. the total energy of the particle, can be determined by the initial conditions of the second-order differential equation from which Eq. (\ref{Jac9}) has been extracted. Now it is easy to solve Eq. (\ref{Jac9}) and to look for the dynamics of the condensate. However, the analysis of the equilibrium point obtained from the extremum of $V(R)$ written as
\begin{eqnarray}
\frac{d V(R)}{dR}& = & 0 \label{Jac11}
\end{eqnarray}
can give some illuminating results. For bright solitons the nonlinear interaction is attractive and the scattering length $a_s < 0$. In this case we shall use $P = -|P|$ and carry out the subsequent analysis by using only the numerical values of $a_s$. We shall make use of Eq. (\ref{Jac11}) to derive a simple physical picture for the collapse dynamics of bright solitons when the trap of the BEC is relaxed in one direction. From Eqs. (\ref{Jac10}) and (\ref{Jac11}) with $P = -|P|$ we get
\begin{eqnarray}
4 \lambda^2 R^4 + |P|R-4& = &0. \label{Jac12}
\end{eqnarray}
The equilibrium point determined by Eq. (\ref{Jac12}) should be a minimum for our system to support a soliton solution and the condition for minimum  ($\frac{d^2V(R)}{dR^2} >  0$) gives
\begin{eqnarray}
4 \lambda^2 R^4 -2 |P|R + 12& = \mu  R^4\,\,, \label{Jac13}
\end{eqnarray}
where $\mu>0$.
Eliminating $|P|$ from Eqs. (\ref{Jac12}) and (\ref{Jac13}) we find that
\begin{eqnarray}
R& = \frac{\displaystyle \sqrt{2}}{\displaystyle (\mu-12\lambda^2)^{1/4}} \label{Jac14}
\end{eqnarray}
is a particular solution of Eqs. (\ref{Jac12}) and (\ref{Jac13}). From Eq. (\ref{Jac14}) and Eq. (\ref{Jac12}) or Eq. (\ref{Jac13}) we get
\begin{eqnarray}
|P|& = \frac{\displaystyle 2\sqrt{2}(\mu-16\lambda^2)}{\displaystyle (\mu-12\lambda^2)^{3/4}}. \label{Jac15}
\end{eqnarray}
The form of equation (\ref{Jac15}) imposes a further restriction on the values of $\mu$ than that given in equation (\ref{Jac14}) and sets a lower bound for it. Using $\mu=\gamma \lambda^2$ we write Eq. (\ref{Jac15}) in the form
\begin{eqnarray}
|P|& = \frac{\displaystyle 2\sqrt{2}(\gamma-16)\lambda^2}{\displaystyle [(\gamma-12)\lambda^2]^{3/4}}. \label{Jac16}
\end{eqnarray}
Thus non-zero values of $P$ will be obtained for $\gamma > 16$ only.

\begin{figure}[!ht]
\begin{center}
\includegraphics[width=0.55\linewidth]{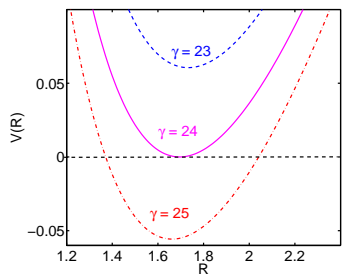}
\end{center}
\caption{The potential $V(R)$ in (\ref{Jac10}) as a function of $R$ for $\gamma = 23, 24,$ and $25$, respectively.}
\label{f1a}
\end{figure}

For $\gamma > 16$ the interaction term vanishes and the corresponding GPE becomes linear and the soliton formation becomes impossible. From Eqs. (\ref{Jac10}) and (\ref{Jac16}) we obtain the expression
\begin{eqnarray}
N& = & \frac{2 \lambda^2}{\pi}\frac{l}{a_s}\frac{(\gamma-16)}{(\gamma \lambda^2-12\lambda^2)^{3/4}} \label{Jac17}
\end{eqnarray}
for the number of atoms that are present in the system. In Fig. \ref{f1a} we plot the potential $V(R)$ in (\ref{Jac10}) as a function of $R$; here we have used the value of $N$ from Eq. (\ref{Jac17}). In this figure we show three curves represented by $V_{23}(R)$, $V_{24}(R)$, and $V_{25}(R)$ corresponding to $\gamma = 23,\,24,$ and $25$, respectively. A common feature of all these potentials is that each of them exhibits a minimum. The curve for $V_{25}$ represents a potential well. The minimum of the well is negative. A mechanical analogy suggests a solution that oscillates between the zeros of $V_{25}$. In this situation the BEC soliton will become unstable and leads to a collapse.

\begin{figure}[!ht]
\begin{center}
\includegraphics[width=0.45\linewidth]{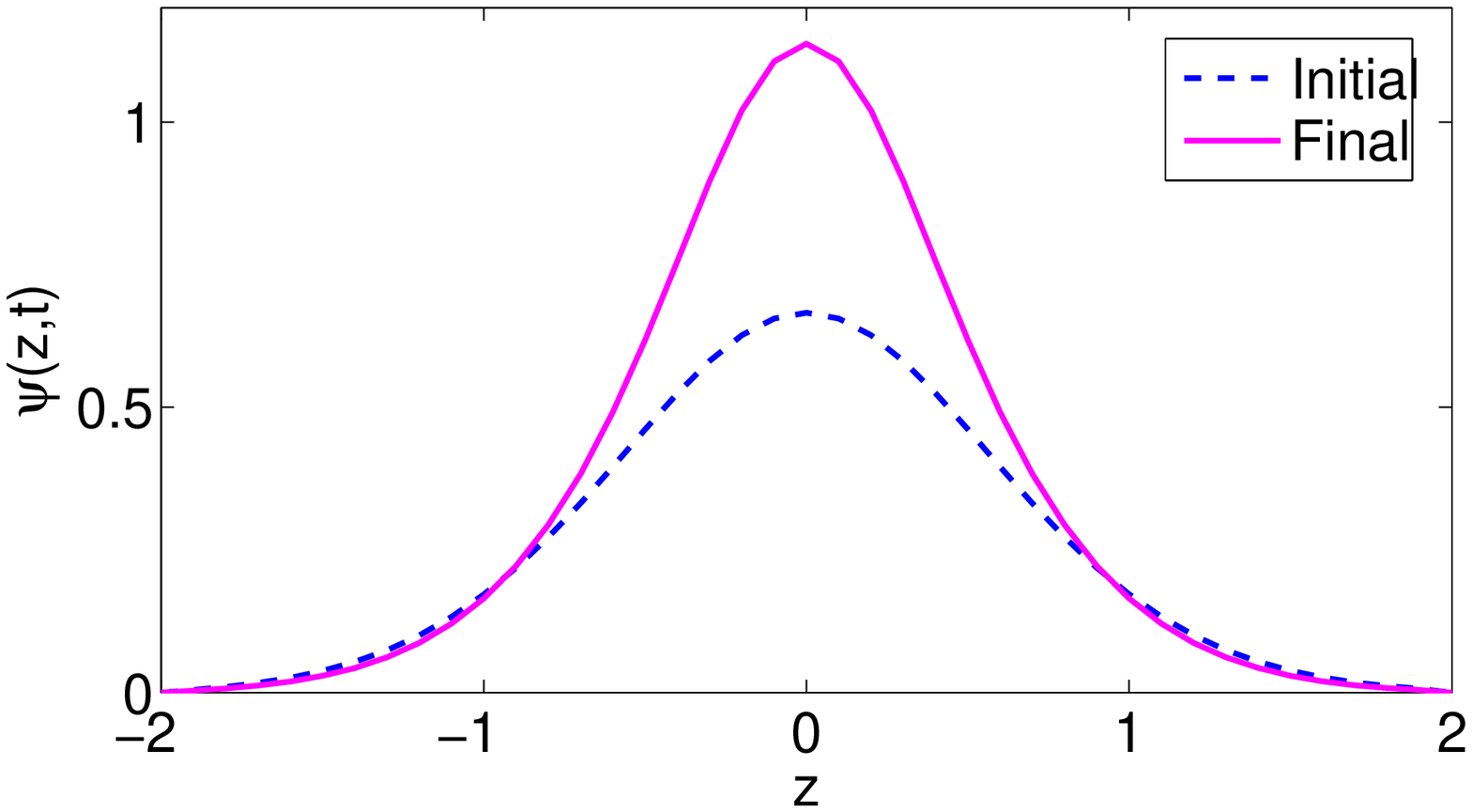}
\vskip 0.6cm
\includegraphics[width=0.4\linewidth]{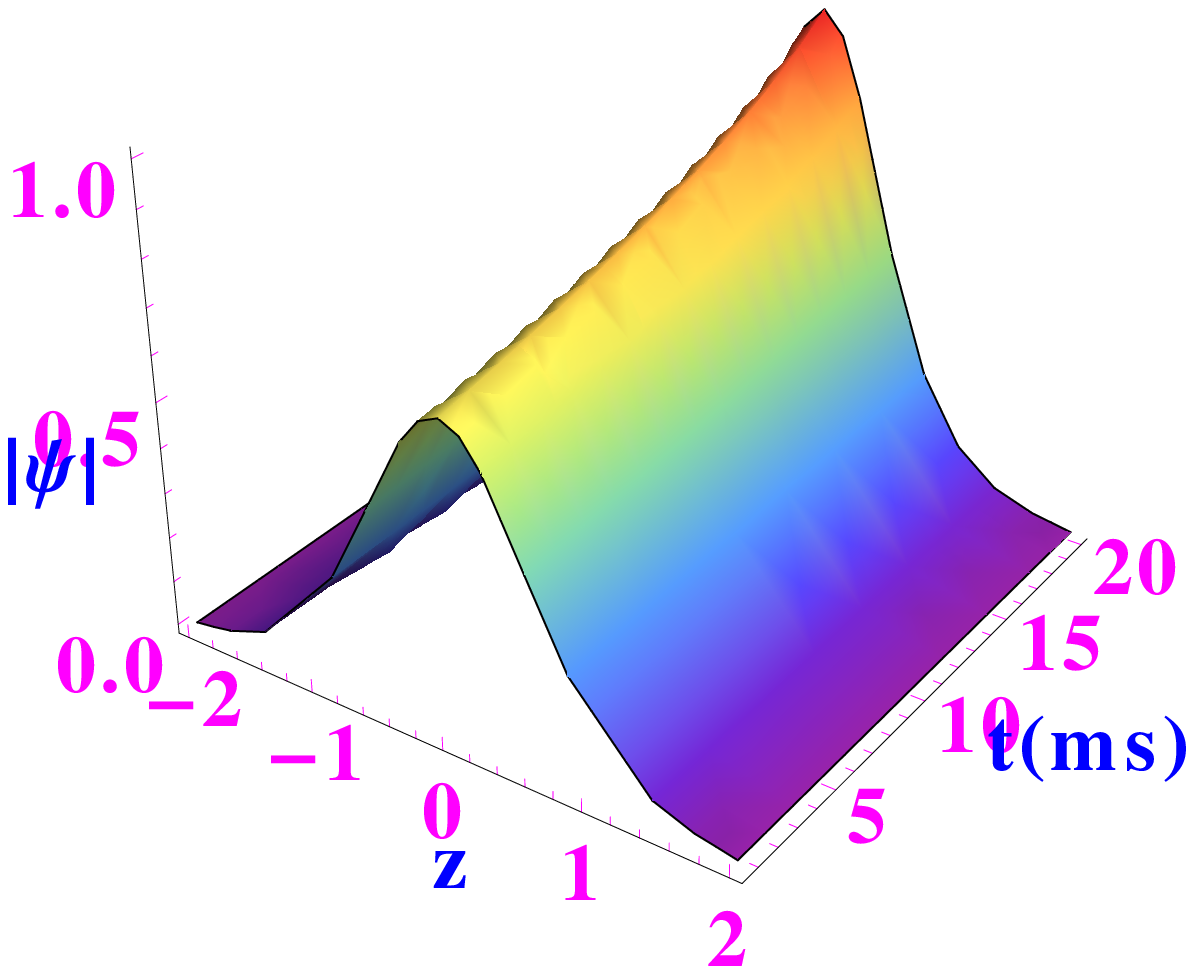}
\hskip 0.06cm
\includegraphics[width=0.4\linewidth]{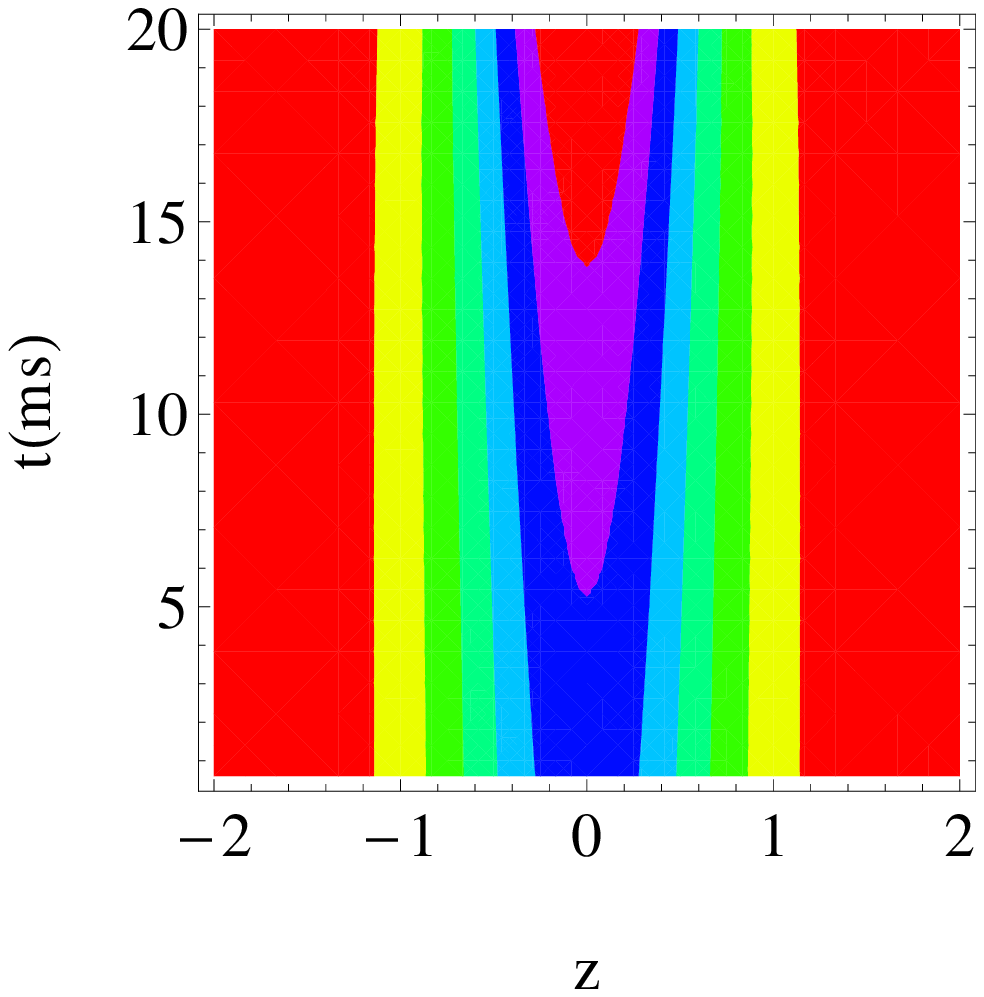}
\end{center}
\caption{Plot of the absolute Gaussian wave function $|\psi|$ illustrating the stability of the condensate below critical number $(< N_c)$. The upper subplot presents the initial (dashed) and final (solid) condensate profiles. In the lower subplots, the left/right-hand-side figure shows the time evolution of the amplitude/contour plot of the condensate for $N (< N_c = 1600)$.}
\label{ffa}
\end{figure}

A similar situation arises for other values of $\gamma > 24$. For $\gamma =24$ the potential well degenerates into a single point such that $V_{24}$ touches the axis at a particular point, where the potential has a stable minimum. Understandably, a particle released at this point will stay there. In the present context this implies that for our chosen value $\gamma=24$ the BEC soliton will be critically stable. The attractive BEC is stable, only when the number of atoms is below some critical number of atoms calculated from Eq. (\ref{Jac17})
\begin{eqnarray}
N_{c}& = & N_{|\gamma=24}=0.7899 \frac{l}{a_s}\sqrt{\lambda}. \label{Jac18}
\end{eqnarray}
For the $^7Li$ atom, mass $m=1.1524 \times 10^{-26}$ Kg, the scattering length $a_s =(-27.6\pm0.5)a_0$ \cite{Bradley1997} and $=(-27.3\pm0.8)a_0$ \cite{Bradley1995}, with Bohr radius $a_0=0.529 \times 10^{-10}$ m, and using the frequencies of the trap in the radial direction $\omega_{\rho}=433$  Hz and axial direction $\omega_z=39$  Hz for the cigar-shaped geometry, and correspondingly $\bar{\omega}=194.095$ and $\lambda=0.20093$, we have calculated the critical number of atoms $(N_c)$= 1667 from Eq. (\ref{Jac18}). Our calculated result is closely matched with the critical value $N_c=1400$ for this geometry reported earlier by the experiment \cite{Bradley1995,Bradley1997,Bradley2} and by the theory \cite{wadati}. The $N_c$ can change with respect to experimental parameters i.e., different trapping frequencies and interaction between the atoms in the condensate are related to the s-wave scattering length tuned by Feshbach resonance technique \cite{FBR}. Due to these factors the initial wavefunction of the condensate is more spread or confined, and such a configuration imposes a more or less severe restriction on the collapse and also change the $N_c$ \cite{wadati}.

\section{Numerical results}
\label{sec4}

We have numerically simulated the stability and collapse dynamics of Q1D BEC that supports our analytical study of the critical number $N_c$. The experimental realization of cigar-shaped attractive BECs has been possible under the strong transverse binding which, in the case of weak or no axial binding, imply one dimensional BECs. We study the stability and the evolution of a collapsing cigar-shaped BEC by solving numerically the time-dependent three-dimensional (3D) GPE (\ref{Jac1}) with strong transverse binding ($\omega_{\rho}=433$  Hz) and less axial trapping ( $\omega_z=39$  Hz), through the SSCN method \cite{Muruganandam2009}. The advantage of the present split-step procedure is that the nonlinear and other linear non-derivative terms can be treated very precisely and this improves the accuracy and stability of the method compared to other methods. We have used spatial grid points $40\times40\times80$; the typical space and time steps for discretization are 0.01 and 0.0001.

\begin{figure}[!ht]
\begin{center}
\includegraphics[width=0.45\linewidth]{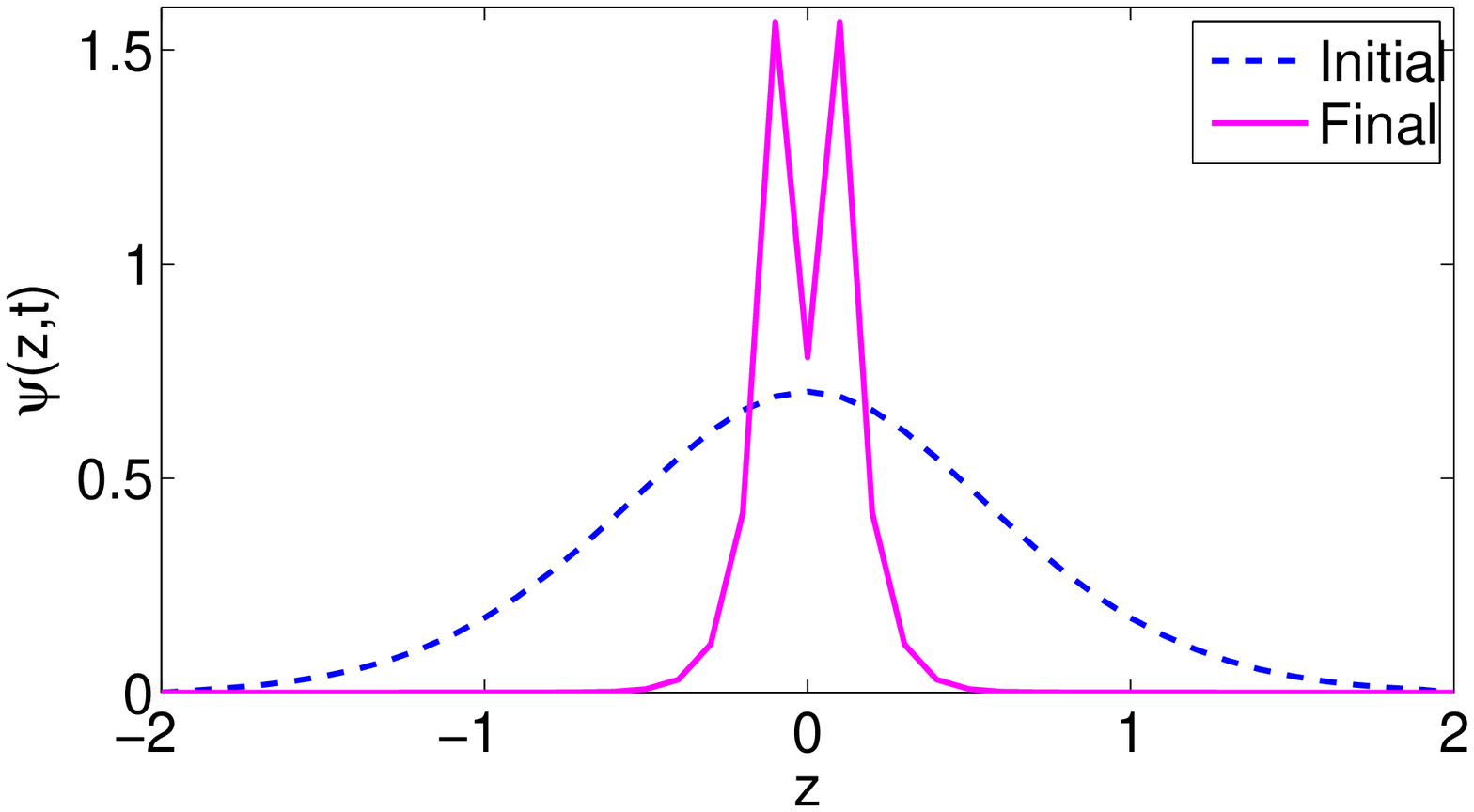}
\vskip 0.6cm
\includegraphics[width=0.4\linewidth]{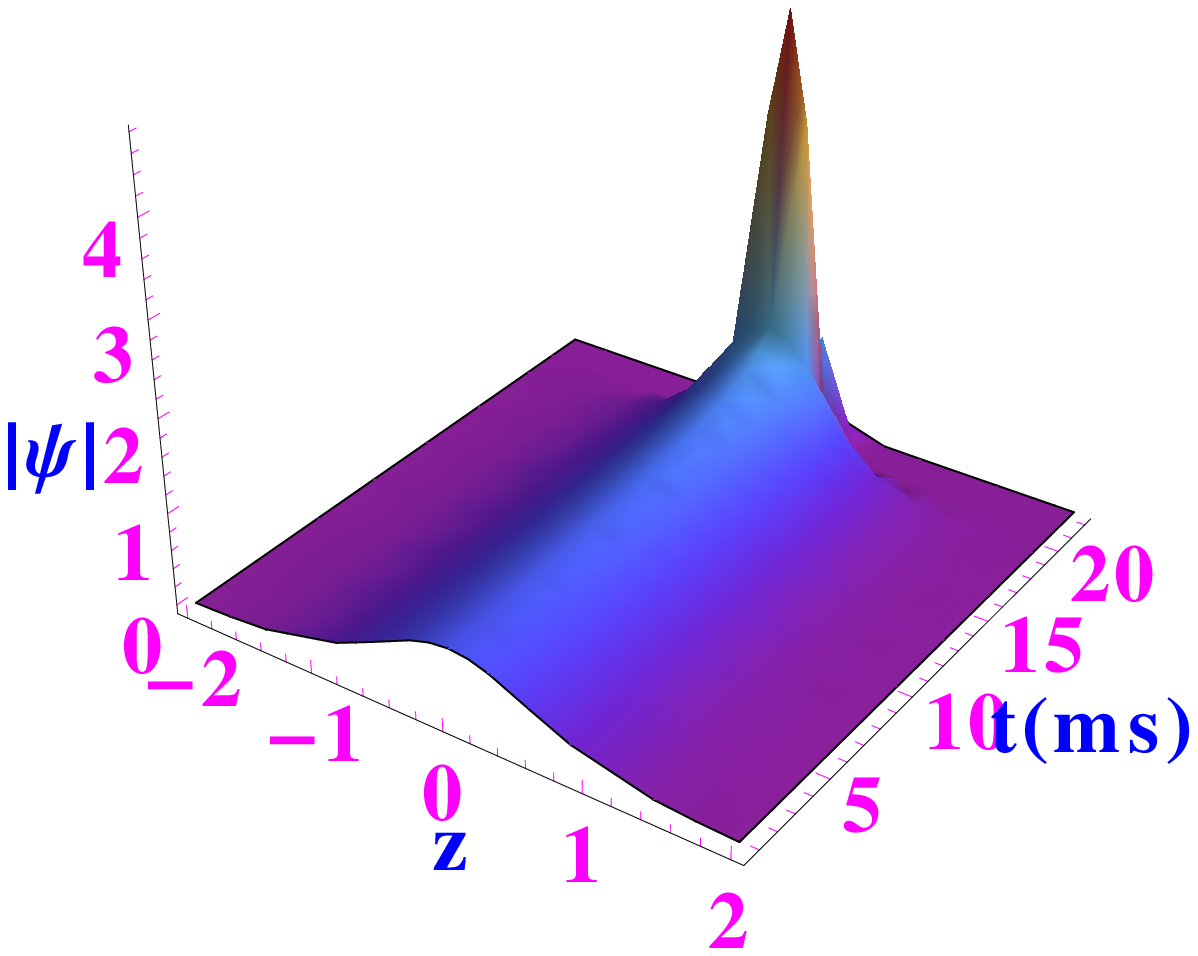}
\hskip 0.06cm
\includegraphics[width=0.4\linewidth]{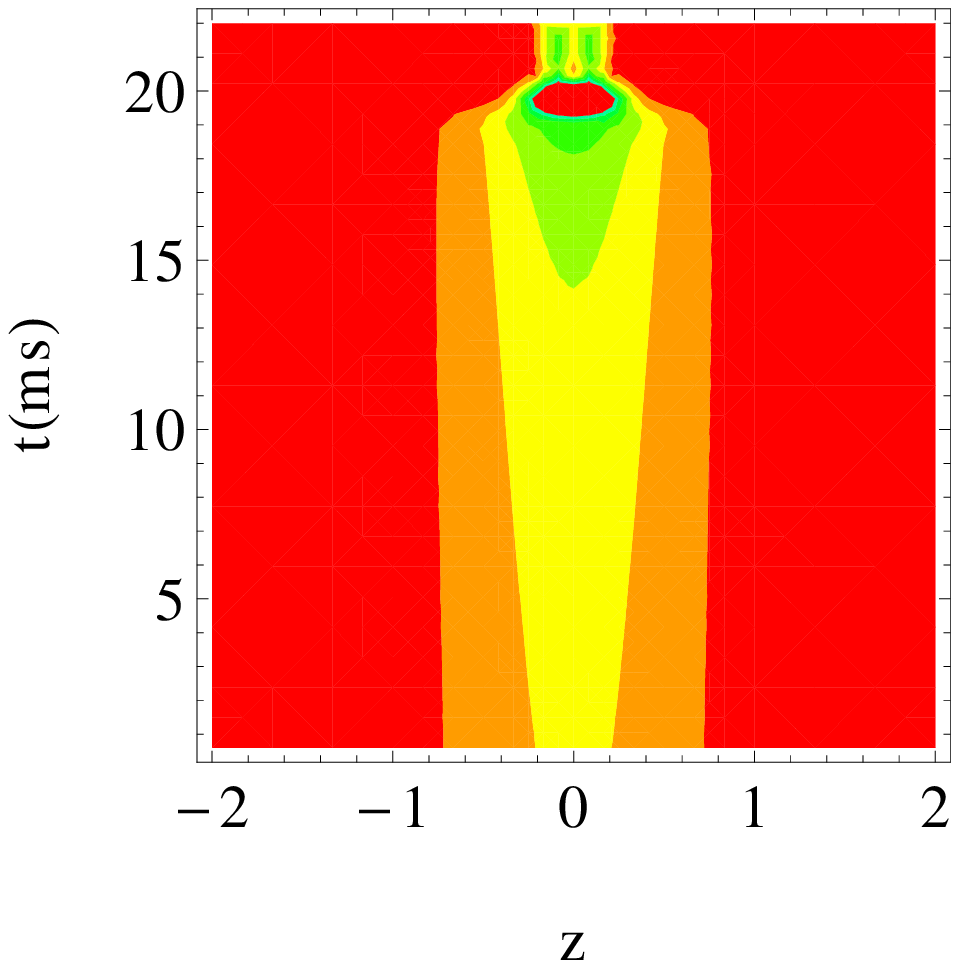}
\end{center}
\caption{Plot of the absolute Gaussian wave function $|\psi|$ illustrating the stability of the condensate at the critical number $(= N_c)$. The upper subplot shows the  initial (dashed) and final (solid) condensate profiles. In the lower subplots the left/right-hand-side figure shows the time evolution of the amplitude/contour plot of the condensate for $N (= N_c ) = 1660$.}
\label{ffb}
\end{figure}

Figure \ref{ffa} depicts the stability of the attractive condensate for the number of atoms present in the system both below $N_c$ and around $N_c$. Here, in the upper plot (see Fig. \ref{ffa}), we have shown that the initial (dashed line) and final (solid line) profiles illustrate the stability of the condensate below the critical number  $N (< N_c)$ = $1600$. In the lower left-hand-side subplot (see Fig. \ref{ffa}), we have presented the time evolution of the amplitude $|\psi|$ of attractive BEC, which shows the gradual shrinking of the width of the condensate with respect to time, but still it is stable for more than $t= 20$ ms. In this case, the system is spatially confined and when the number of BEC atoms is below a certain critical value $N_c$, the zero-point motion of the atoms serves as a kinetic obstacle against collapse, which allows to form a metastable BEC.

\begin{figure}[!ht]
\begin{center}
\includegraphics[width=0.45\linewidth]{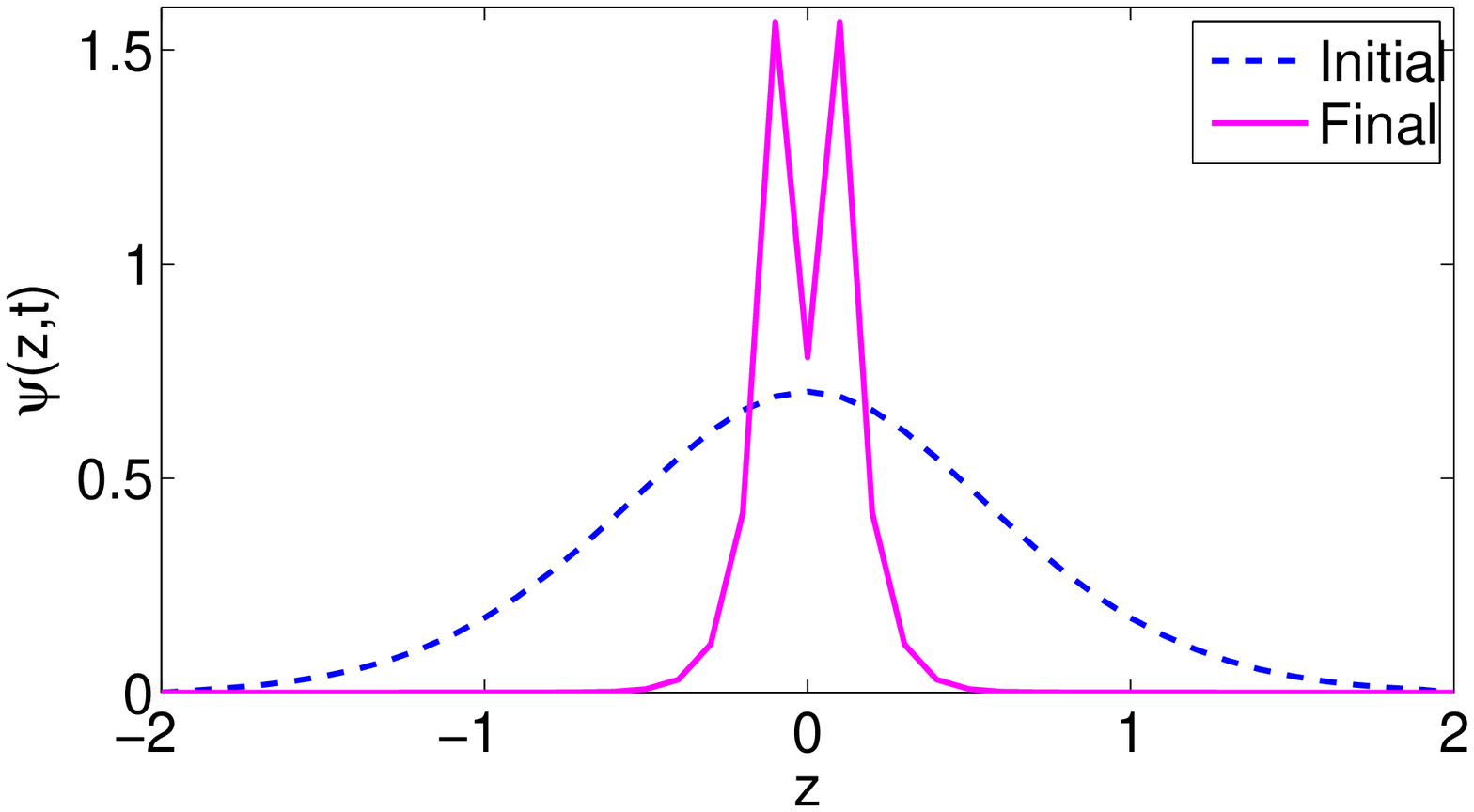}
\vskip 0.6cm
\includegraphics[width=0.4\linewidth]{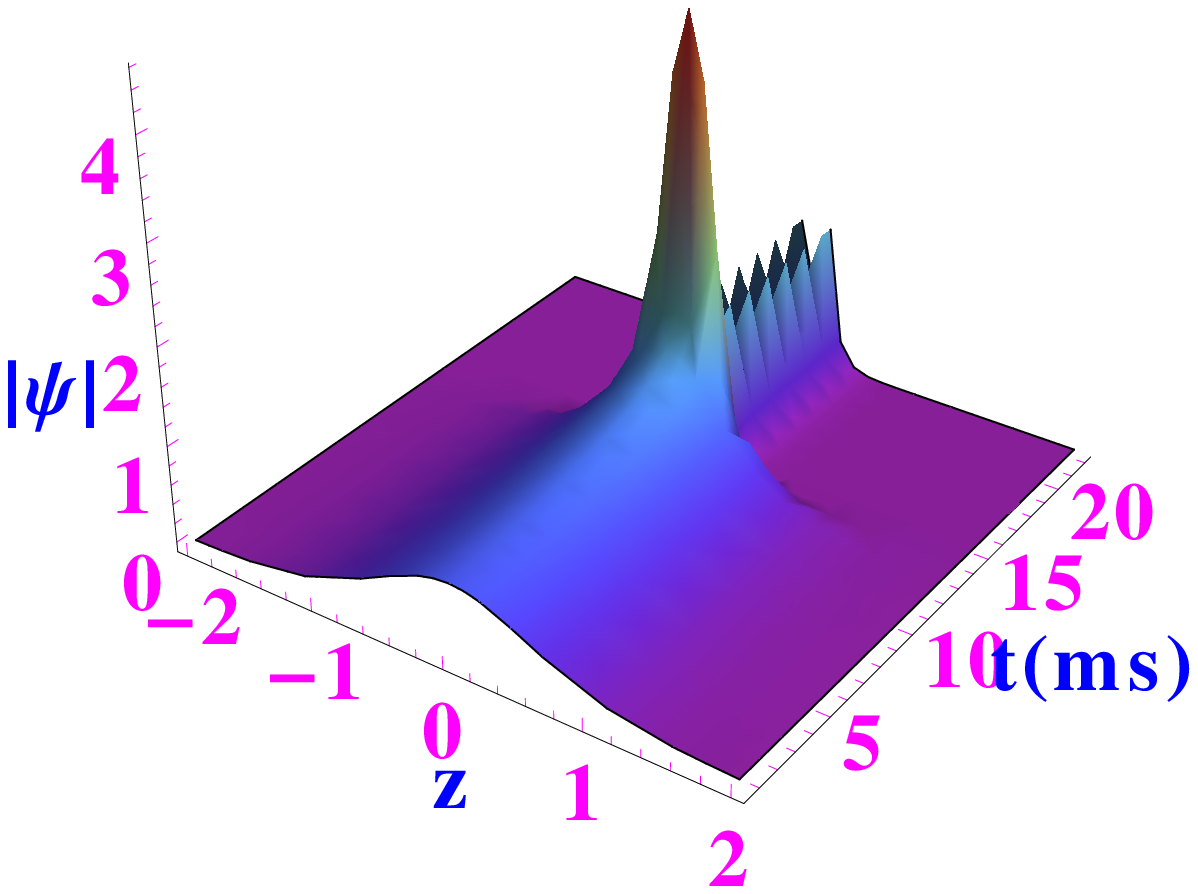}
\hskip 0.06cm
\includegraphics[width=0.4\linewidth]{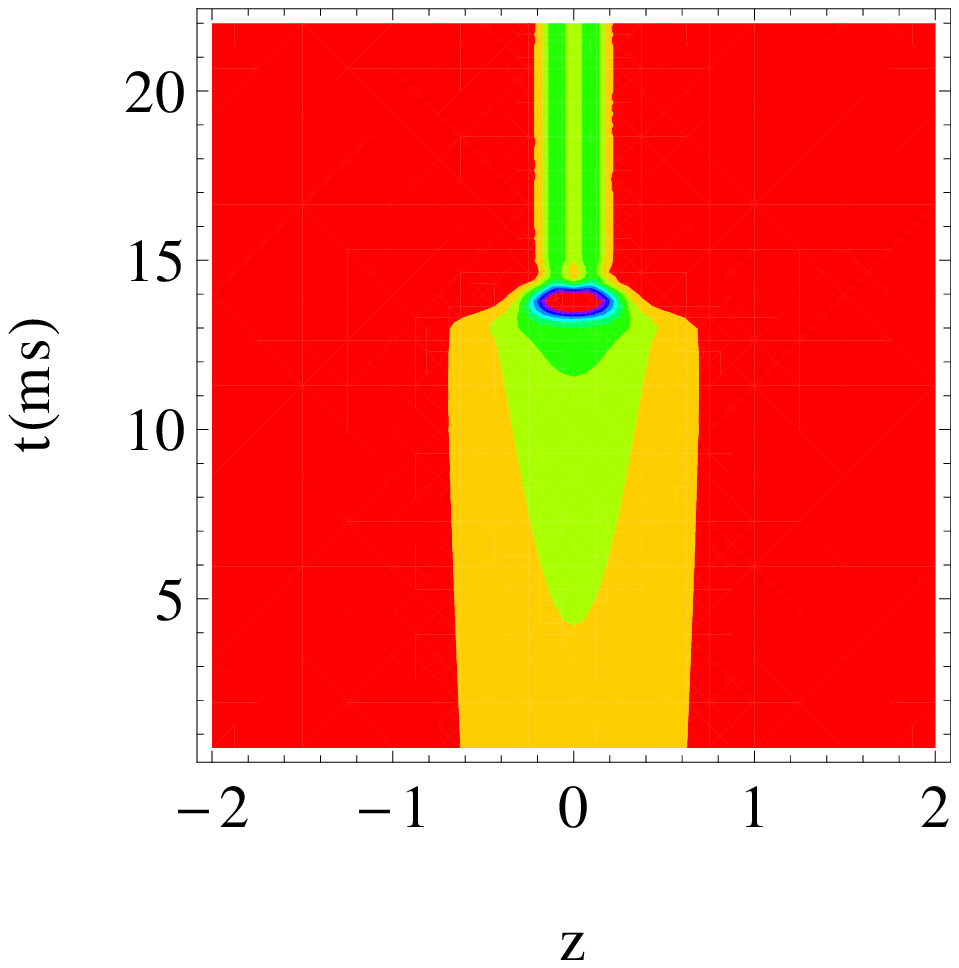}
\end{center}
\caption{Plot of the absolute Gaussian wave function $|\psi|$ illustrating the stability of the condensate above critical number $(> N_c)$. The upper subplot depicts the initial (dashed) and final (solid) profile of the condensate. In the lower subplots, the left/right-hand-side figure shows time evolution of the amplitude/contour plot of the condensate for $N (> N_c) = 1700$.}
\label{ffc}
\end{figure}

\begin{figure}[!ht]
\begin{center}
\includegraphics[width=0.45\linewidth]{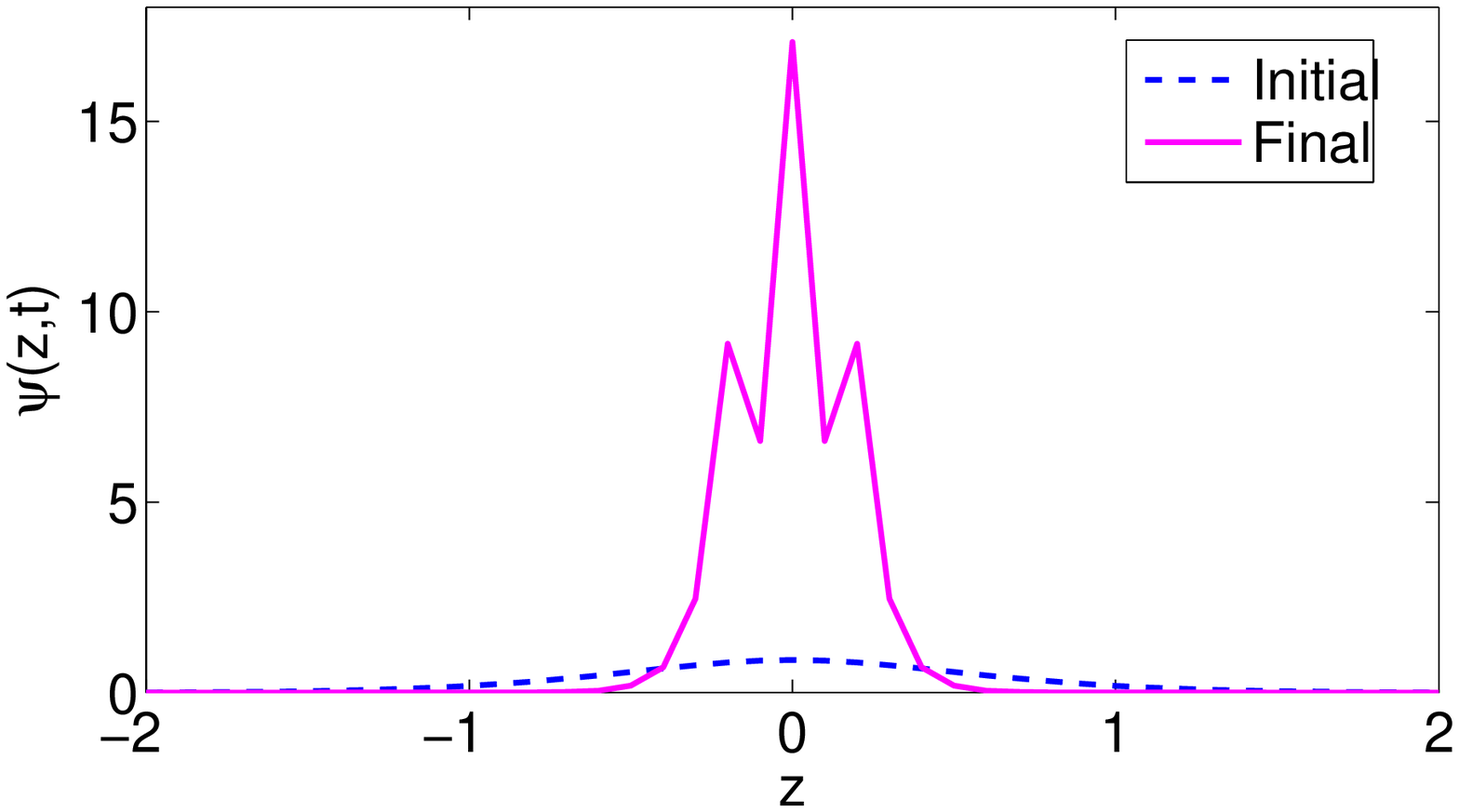}
\vskip 0.6cm
\includegraphics[width=0.4\linewidth]{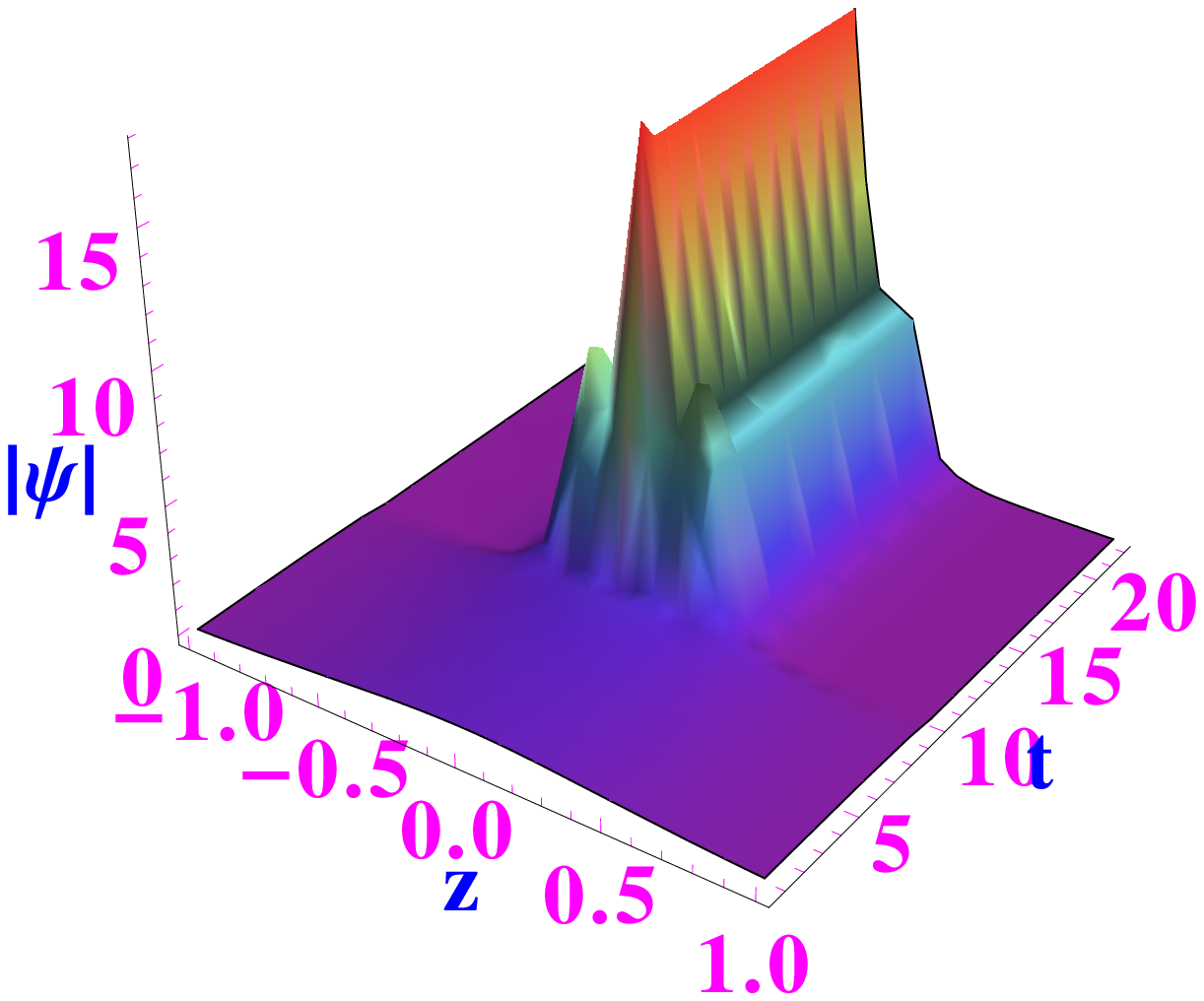}
\hskip 0.06cm
\includegraphics[width=0.4\linewidth]{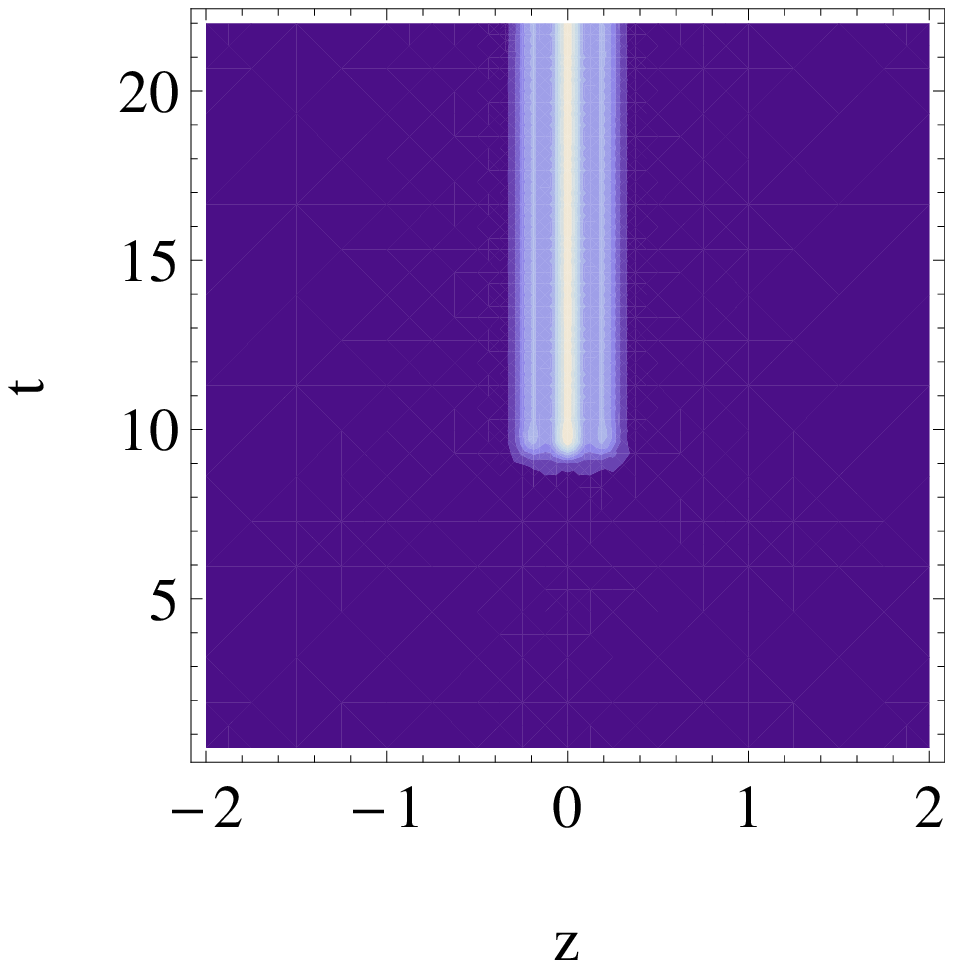}
\end{center}
\caption{Plot of the absolute Gaussian wave function $|\psi|$ illustrating the stability of the condensate far above the critical number $(> N_c)$. The upper subplot depicts the initial (dashed) and final (solid) profiles of the condensate. In the lower subplots, the left/right-hand-side figure shows the time evolution of the amplitude/contour plot of the condensate for $N (>N_c) = 2000$.}
\label{ffd}
\end{figure}

But, in Fig. \ref{ffb}, we have shown the sudden explosion of attractive BEC followed by an implosion during the evolution after $t= 20$ ms for $N=1660$.  Due to high attractive interaction of the BEC atoms the condensate implodes and then a sudden explosion occurs because locally near the center of BEC where the atomic density exceeds a certain critical value, the centripetal force then weakens, and the atoms that gathered in this narrow central region are ejected due to the 'quantum pressure' arising from the uncertainty principle. Here, interestingly, we have shown an explosion followed by an implosion. Further,  when we increase the number of atoms in the condensates, say, $N=1700$, the condensate collapses earlier at $t= 14 ms$ and explodes; this phenomenon is clearly shown in Fig. \ref{ffc}. Then we have considered up to $N=2000$ atoms and we have shown in Fig. \ref{ffd} another type of explosion pattern.

\begin{figure}[!ht]
\begin{center}
\includegraphics[width=0.25\linewidth]{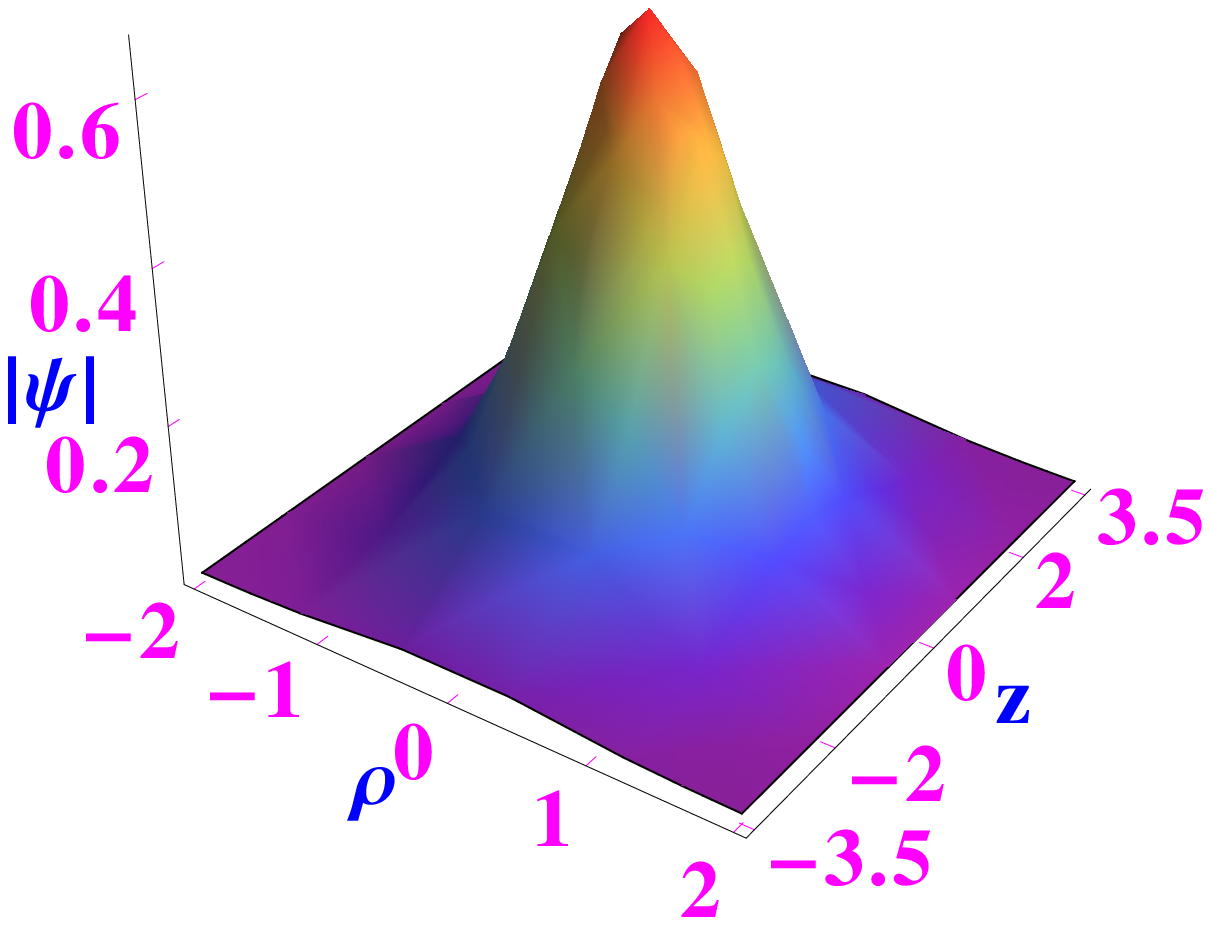}
\includegraphics[width=0.25\linewidth]{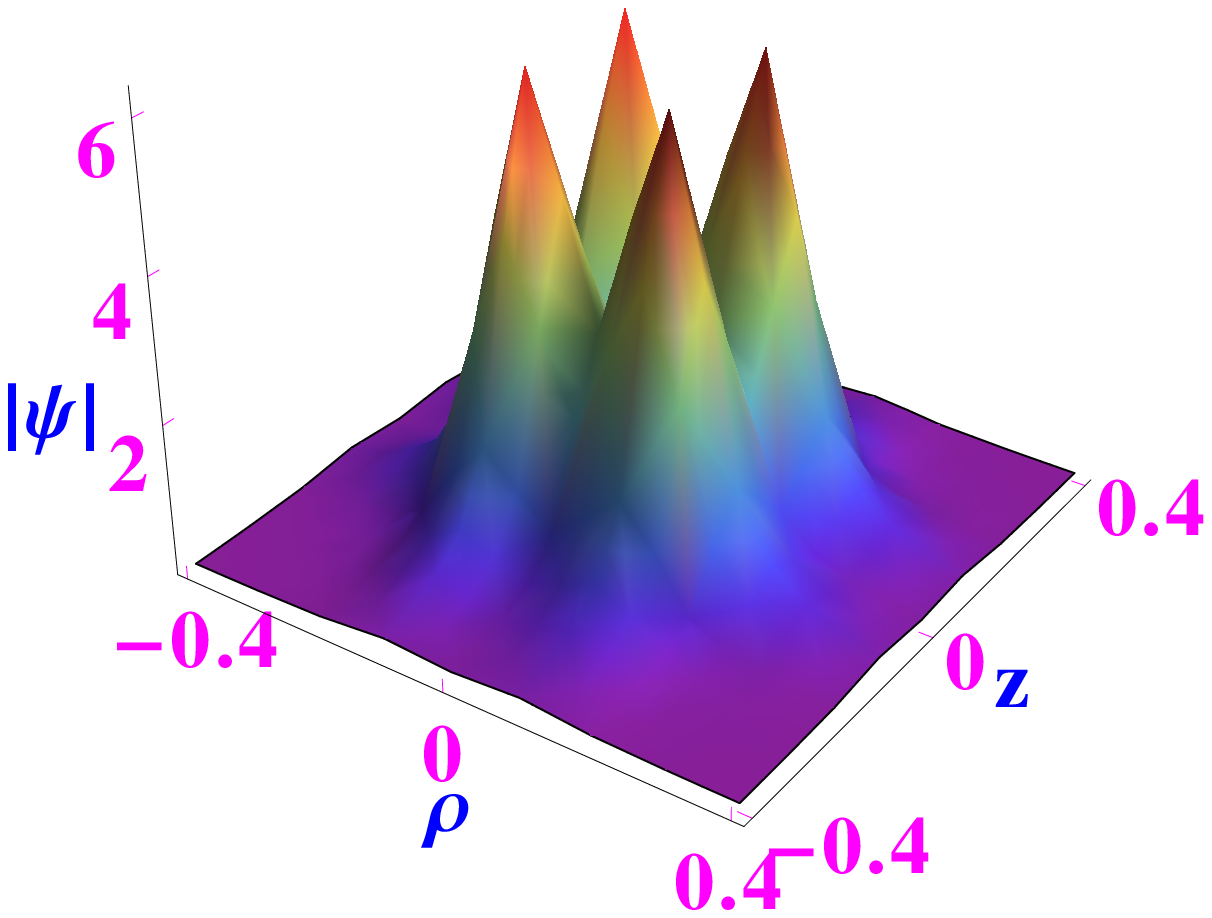}
\includegraphics[width=0.25\linewidth]{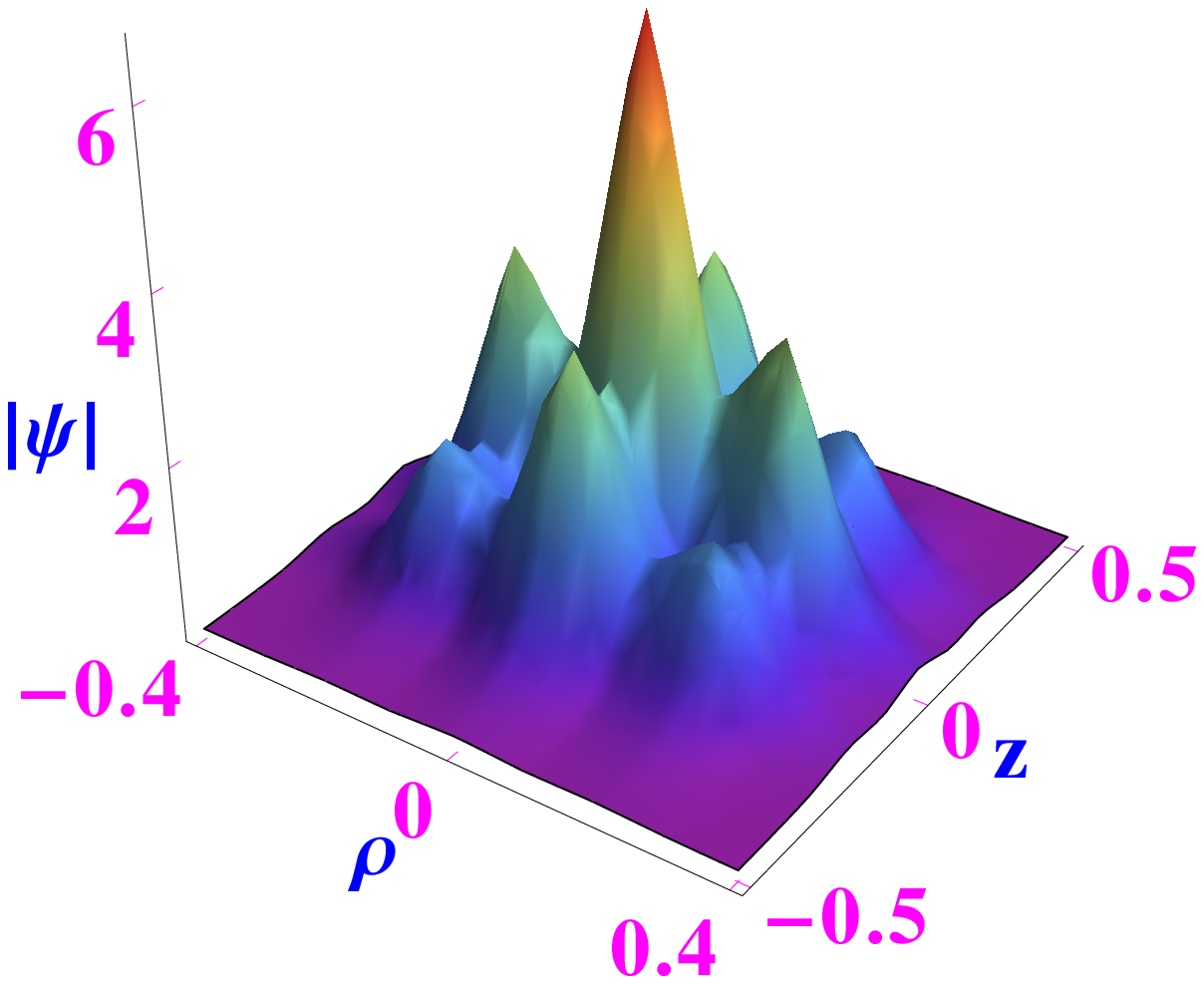}
\includegraphics[width=0.25\linewidth]{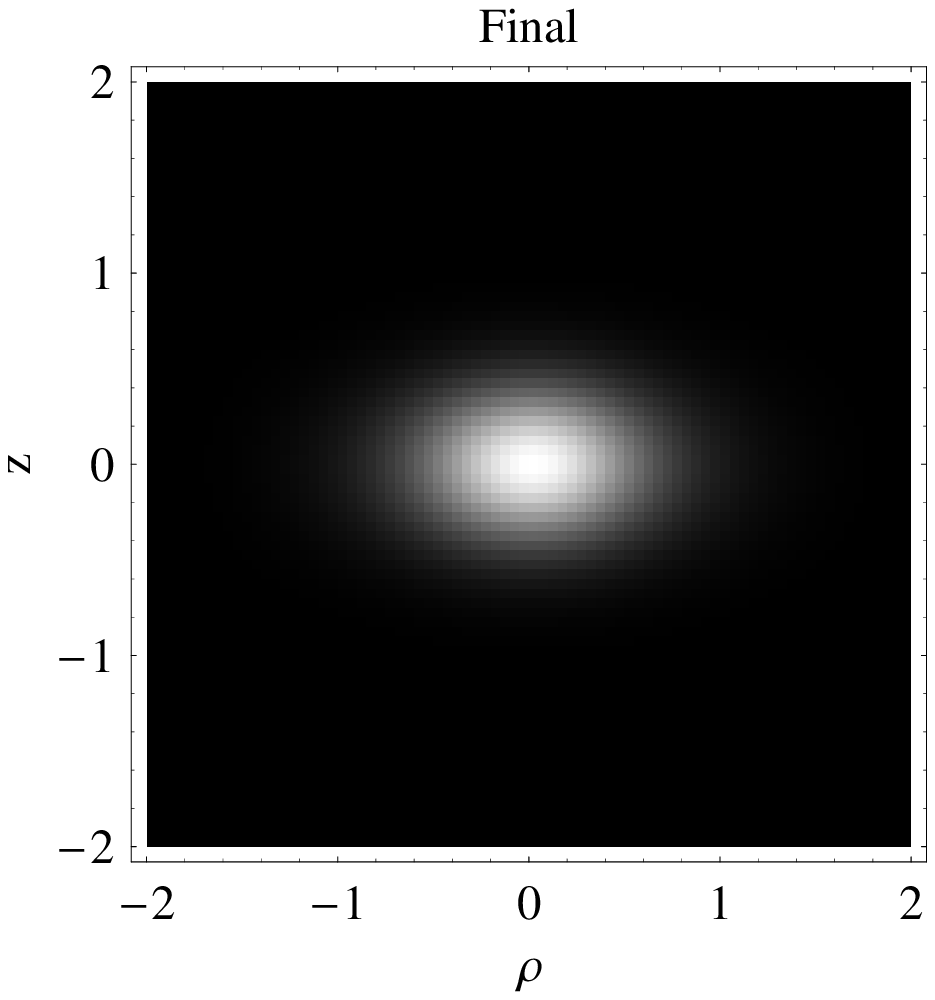}
\includegraphics[width=0.25\linewidth]{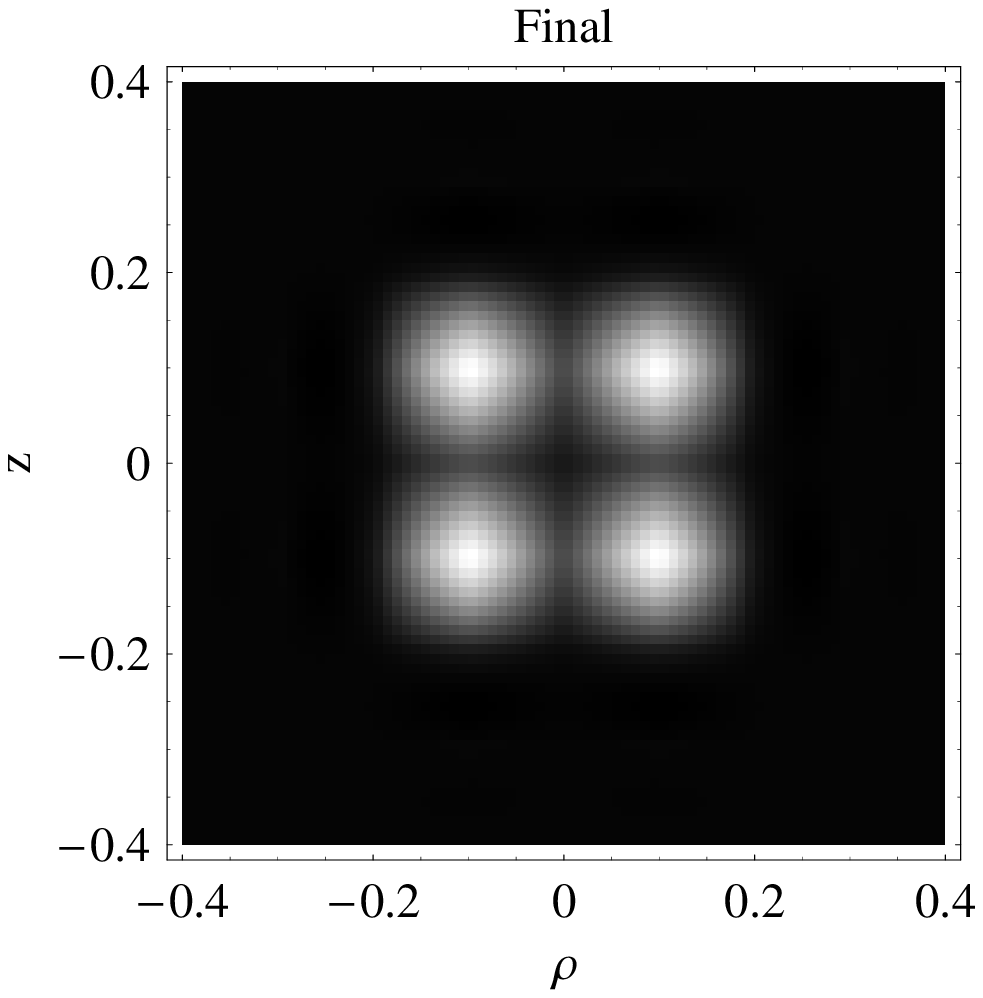}
\includegraphics[width=0.25\linewidth]{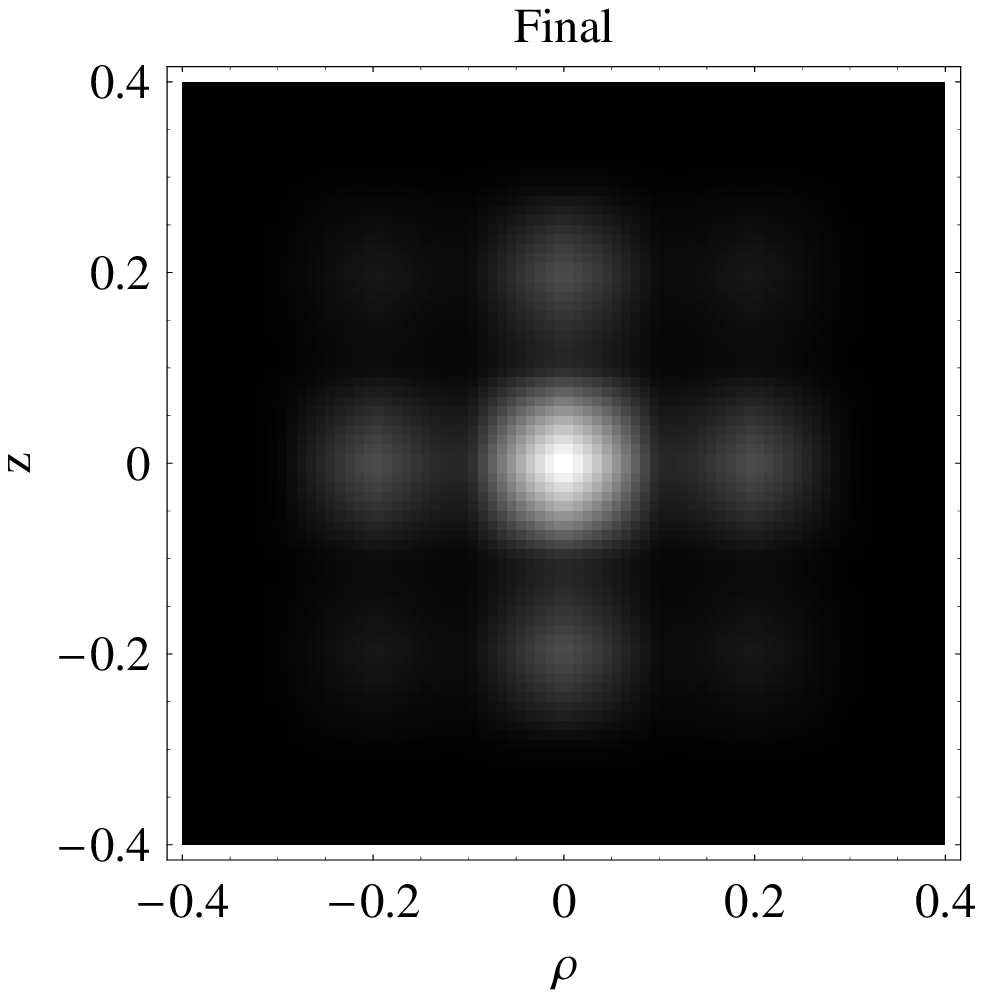}
\end{center}
\caption{The amplitude (upper)/contour (lower) plots correspond to the final condensate for $N=$ (left$\rightarrow$ $1600$), (middle$\rightarrow$ $1660$ and $1700$) and (right$\rightarrow$ $2000$), respectively.}
\label{ffg}
\end{figure}

We have shown in Fig. \ref{ffg} the amplitude/contour plots corresponding to the final condensate for $N=$ (left$\rightarrow$ $1600$), (middle$\rightarrow$ $1660$ and $1700$) and (right$\rightarrow$ $2000$), respectively. Below the critical number of atoms we can see that the condensates are in the metastable states. As we increase the number of the atoms to critical value,  the condensate is shrinking with time and explodes into four major peaks followed by its implosion. If we increase the number of atoms even more, another type of explosion phenomenon occurs giving rise to a different pattern containing a central high density peak surrounded by small density peaks (see the right panel in Fig. \ref{ffg}).

\section{Conclusion}
\label{sec5}

In conclusion, we have shown theoretically the possibility of a subsequent explosion after implosion of an attractive BEC by solving numerically the time-dependent GPE (\ref{Jac1}) through the SSCN method. We have calculated  the  critical number of atoms through the VA method and investigated the stability of the condensate by numerical techniques. The condensate collapses when the number of atoms exceeds the critical value. Beyond $N_c$, the attractive BEC first exhibits explosion followed by implosion and then it exhibits a subsequent explosion. Our results are in good agreement to those obtained in the Bosenova experiment at JILA \cite{Donley2001} and at Rice University \cite{Bradley1997,HU2000} that the condensates with a particle number exceeding the critical value are unstable against collapse. %The novelty of our work is new kind of structure formation during the explosion of the attractive BEC followed by an implosion.

\vskip 0.5 cm

{\bf Acknowledgement}

S.S. thanks University Grants Commission (UGC) for offering support through a Post-Doc under the Dr. D.S. Kothari Post Doctoral Fellowship Scheme. K.P. thanks to the DST-DFG, DST, CSIR and UGC, Government of India, for the financial support through major projects.

\end{document}